\documentstyle[psfig]{mn}
\newif\ifAMStwofonts

\newcommand{\microJy}{$\mu\mbox{Jy}$}

\title[Radio sources in 47 Tuc.]
	{Radio sources near the core of globular cluster 47 Tucanae}
\author[D. McConnell and J.G. Ables]
	{D. McConnell$^1$ and J.G.Ables$^2$\\
$^1$ATNF, CSIRO, Locked Bag 194, Narrabri NSW 2390, Australia\\
$^2$Telecommunications and Industrial Physics, CSIRO, PO Box 76, Epping, NSW
2121, Australia\\
}
\date{Version: \today}

\pagerange{\pageref{firstpage}--\pageref{lastpage}}
\begin{document}

\maketitle

\label{firstpage}

\begin{abstract}
We present ATCA radio images of the globular cluster 47 Tucanae made at 1.4 and
1.7~GHz and provide an analysis of the radio sources detected within 5\arcmin\ of the
cluster centre. 11 sources are detected, most of which are clustered about the
core of 47 Tuc. Both of the pulsars in 47 Tuc whose positions are known
can be identified with sources in the 1.4~GHz image. The source distribution
has a characteristic radius of $\sim$100\arcsec, larger than the 23\arcsec\ radius of the
cluster core. We compare source positions with the positions of nine X-ray
sources and find no correspondence.

\end{abstract}

\begin{keywords}
globular clusters: 47 Tucanae -- pulsars
\end{keywords}

\section{Introduction}

The globular cluster 47 Tucanae is known to contain many short period
(millisecond) radio pulsars. Robinson et al.\ \shortcite{rlm+95} reported 11
pulsars observed with the Parkes radiotelescope at 640 MHz and 436 MHz.
Millisecond pulsars as a class are intrinsically faint. Coupled with the
relatively great distance to 47 Tuc (4.6 kpc), this means that most of the
pulsars are difficult to detect. Robinson et al. report that many detections
rely on the occasional flux enhancements produced by the focussing effects of
diffraction and refraction in the irregular interstellar medium. Many of the
pulsars show period variations characteristic of motion in a binary orbit. The
use of pulse timing to determine pulsar positions in the cluster has been made
difficult by both their intermittent detectability and the extra complexity of
timing analyses arising from their orbital motions. In 1995, Robinson et al.
were able to report positions for only two of the pulsars --- 47 Tuc C and 47
Tuc D.

The positions of the pulsars relative to the cluster centre are interesting
because of the relation to the dynamics and evolutionary history of the cluster
core. For example, pulsars C and D have a negative period derivative which is
attributed not to an intrinsic increase in rotation rate, but to their
accelerated motion in the cluster's gravitational field.

Here we report radio images of 47 Tuc made at 1.4~GHz and 1.7~GHz with the
Australia Telescope Compact Array (ATCA) and an
attempt to seek point radio sources which may be identified with pulsars in the
cluster. At 1.4~GHz, the sky away from the Galactic plane is dominated by
extragalactic sources with small angular sizes. The number density of these
sources increases, naturally, with decreasing flux density and at flux levels
expected of the radio pulsars in 47 Tuc, tens of sources are expected in each
ATCA primary beam. Several reports of studies of 1.4~GHz source densities are
available as a guide to the expected number of background sources. In
particular, the ``Phoenix Deep Survey'' \cite{hmcr98} was conducted
at 1.4~GHz with the ATCA and provides a useful reference. In the analysis of
these observations of 47 Tuc we study three source characteristics which might
distinguish the radio pulsars from the background sources. (1) Spectral index:
the spectral index $\alpha$ (where $S \sim \nu^{\alpha}$) of pulsar emission
tends to be less than ($\alpha < -1.2$) the typical value for extragalactic sources for
which $\alpha \simeq -0.8$. (2) Flux variability: the very small angular size of radio pulsars
leads to large flux variations due to scintillation in the interstellar medium.
(3) Spatial distribution: background sources are distributed uniformly whereas
the pulsars are expected near the cluster core.

Radio images of other globular clusters have been reported by Kulkarni et al.
\shortcite{kgwm90}. They observed five clusters with the VLA and detected the
pulsars in clusters M4 and M28, one of now eight known pulsars in M15, a
single unidentified radio source in M3 and no radio sources in M92.

\section{Observations and data reduction}

The Australia Telescope Compact Array (ATCA) was used to observe 47 Tuc at
1.408~GHz and 1.708~GHz. At each frequency a 128 MHz bandwidth was used, each
band split into 64 spectral channels. The cluster was observed on 1992 Jan 24
in the 6D array configuration and on 1992 Apr 23,24,25 in the 6C configuration.
Four fields close to the cluster centre were observed. Table 1 gives the field
centres and their times of observation.

\begin{table*}
 \caption{Summary of observations. Six images were made at each frequency, five from each
12-hour integration and a sixth, the mosaic, as a combination of I--V. The sensitivity of
each observation and the mosaic is indicated by the root mean square value of pixel
brightness ($\sigma$), expressed as \microJy/beam, and measured over the central
10\arcmin\ of the cluster.}

\begin{tabular}{cllllrr}
\hline
Observation & \multicolumn{1}{c}{$\alpha$} & \multicolumn{1}{c}{$\delta$} &
\multicolumn{1}{c}{Date} & \multicolumn{1}{c}{Time} &
\multicolumn{1}{c}{$\sigma_{1.4}$}
& \multicolumn{1}{c}{$\sigma_{1.7}$}\\
      & \multicolumn{1}{c}{(J2000)} & \multicolumn{1}{c}{(J2000)} &
       \multicolumn{1}{c}{(UT)} & \multicolumn{1}{c}{(UT)}
       &\multicolumn{1}{c}{(\microJy)} &\multicolumn{1}{c}{(\microJy)}\\
\hline
I   & $00:24:06$ & $-72:05:00$ & 1992 Jan 24    & 06:00 - 11:05 & 66  & 75  \\
    &            &             & 1992 Jan 24-25 & 22:33 - 04:48 &     &     \\
II  & $00:24:06$ & $-72:05:00$ & 1992 Apr 23-24 & 16:40 - 02:40 & 74  & 71  \\
III & $00:25:12$ & $-71:51:00$ & 1992 Apr 24-25 & 14:41 - 02:35 & 102 & 130 \\
IV  & $00:26:06$ & $-72:17:00$ & 1992 Apr 24    & 03:38 - 14:22 & 133 & 186 \\
V   & $00:20:54$ & $-72:07:00$ & 1992 Apr 25    & 02:45 - 15:39 & 118 & 162 \\
mosaic &         &             &                &               &  42 & 46  \\
 \end{tabular}
\end{table*}

The central field was observed in two separate 12-hour integrations which are
listed individually as I and II in the table. Fields III IV and V were offset
from the cluster centre by 15\arcmin, less than the $\sim$17\arcmin\ to the
half-sensitivity point of the ATCA primary beam at 1.4~GHz. Thus all five
integrations individually had good sensitivity to emission from the cluster
centre, and in combination provided a sensitive image over a 35\arcmin\ region
centred on the cluster.

The {\sc miriad} data reduction package was used to calibrate the data and form
images. The antenna gains were calibrated using brief observations of the
source B2353-686 ($S_{1.4} =  1.05$ Jy, $S_{1.7} = 1.00$ Jy) and the flux scale
was referred to the ATCA primary flux calibrator B1934-638 ($S_{1.4} =
14.9$~Jy; $S_{1.7} = 14.0$~Jy). Images were formed for each integration at both
frequencies with reference position (tangent point) at $\alpha$(2000) =
00:24:05\fs83, $\delta$(2000) = -72:04:51\farcs4, the nominal cluster centre
\cite{gys+92}. Visibility data from each 2 MHz spectral
channel were gridded separately allowing the full primary beam to be imaged
with very little bandwidth smearing. An imaging cell size of 1\farcs5 was used
to provide adequate sampling of the synthesized beam. The image for each
integration was made 2041 pixels square (51\arcmin). The final image at each
observing frequency was formed as a linear combination of the images from each
field. In the remainder of this paper we discuss a central region of the image
of diameter 10\arcmin\ and centred on the cluster.

\section{Image Analysis and results}
\subsection{The image noise statistics}
Five sub-images, corresponding to the integrations listed in Table 1, were made
at each frequency. A sixth, the final (mosaic) image, was formed by linear
combination of the five. To allow direct comparison, the five sub-images were
corrected for the sensitivity variation across the primary beam. Table 1 gives
the noise in each sub-image in the central 10\arcmin\ of the cluster as the
root-mean-square of their pixel values ($\sigma$). The higher noise levels in
sub-images IV and V are possibly due the effects of imperfect antenna pointing
during the integration and the presence of relatively strong sources (92~mJy
and 41~mJy) near the halfpower levels of those fields. The size (FWHM) of the
synthesized beam in the combined images was $5\farcs7 \times 5\farcs2$ at
1.4~GHz and $4\farcs5 \times 4\farcs0$ at 1.7~GHz.

\subsection{Radio sources}
Examination of the images shows a number of point sources. Sources within a
5\arcmin\ radius of the cluster centre were analysed. All sources with a peak
brightness exceeding 4.5$\sigma$ in the 1.4~GHz mosaic were selected. Their
positions and fluxes were estimated by fitting each with a gaussian function
with major and minor axes equal to those of the synthesized beam. Because of
the known flux variability of pulsars in the cluster, each of the sub-images (I
- V) were searched for sources coincident in two sub-images but that were not
strong or persistent enough to rise above the detection threshhold in the
mosaic. No such sources were found. All sources detected at 1.7~GHz were also
visible at 1.4~GHz. Table 2 lists the sources detected in the 1.4~GHz
image. The 1.7~GHz image was analysed in the same way and 1.7~GHz fluxes are
reported for each source in Table 2. Two sources (2, 4) had a 1.7~GHz
brightness below the 4.5$\sigma$, threshhold, but for each a successful fit to
a model gaussian source was made, yielding 1.7~GHz fluxes of 165\microJy and
177\microJy (3.6$\sigma$ and 3.8$\sigma$) respectively. Two further sources (9, 11) had no
detectable 1.7~GHz emission and upper limits corresponding to 3$\sigma$ are quoted in
Table 2. The radial distance of each source from the cluster centre
($\alpha$(2000) = 00:24:05\fs83, $\delta$(2000) = -72:04:51\farcs4, Guhathakurta et
al., 1992\footnote{Note that an independent determination of the cluster centre \cite{cmps93}
differs by 1\farcs4 which is not significant in this analysis.}) is listed in the
table. Table 2 also lists, in column 8, the spectral index of each source.
Fig. 1 shows the location and relative flux density of the 1.4~GHz sources
superposed on an optical image (from the Digital Sky Survey) of the cluster.

\begin{figure*}
\psfig{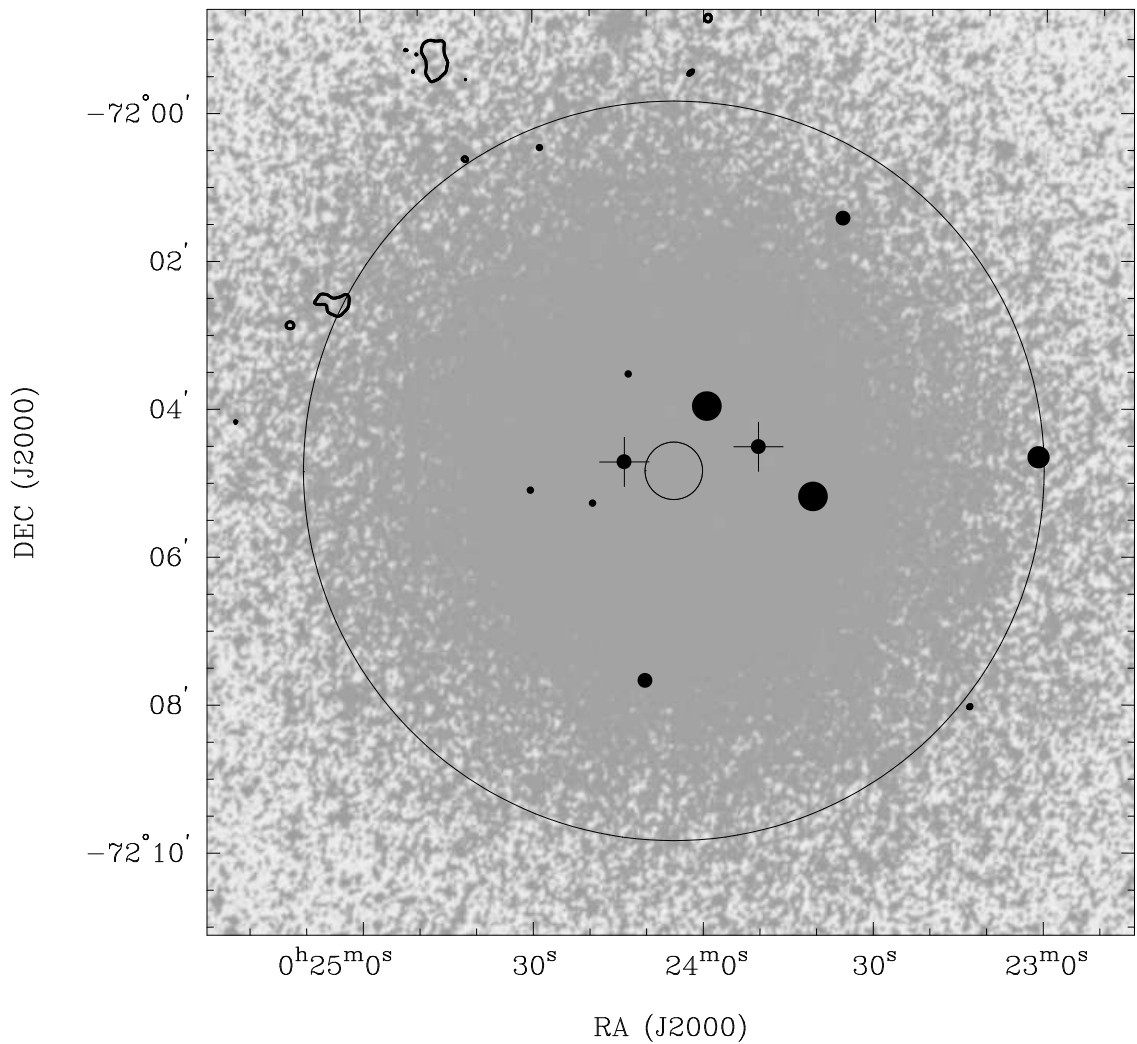}
\caption{
The positions of sources detected in the 1.4~GHz image of 47 Tucanae
shown against an optical image (from the Digital Sky Survey) of the
cluster. The large circle
marks the boundary of the region analysed. Each point source is marked
as a filled circle with size proportional to the logarithm of the
source flux S$_{1.4}$. The faintest sources indicated have S$_{1.4} >
190$\microJy. Two extended sources of emission are indicated near
$\alpha$(2000) = 00:24:47, $\delta$(2000) = -71:59:15 and
$\alpha$(2000) = 00:25:04, $\delta$(2000) = -72:02:36 by a contour at
220\microJy/beam. These regions have integrated 1.4~GHz fluxes of
approximately 60~mJy and 6.5~mJy respectively. The small central
circle marks the position and size ($r_{c} \simeq$ 23\arcsec) of the
cluster core. The locations of pulsars C and D as determined from
pulse timing are indicated by crosses (pulsar C to the west).
}
\end{figure*}

\begin{table*}
\caption{
Point sources within 5\arcmin\ of the cluster centre. Sources with
positions corresponding to known timing positions of pulsars are
indicated by the pulsar letter designation in column 2. Column 5 gives
the distance of each source from the cluster centre $\alpha$(J2000) =
00:24:05\fs83, $\delta$(J2000) = -72:04:51\farcs4. The uncertainties
given for source positions (columns 3, 4) relate to the last digit or
digits quoted and were determined by the technique described in the
text.
}

\begin{tabular}{ccllrrrl}
\hline
\multicolumn{1}{c}{(1)} & \multicolumn{1}{c}{(2)} & \multicolumn{1}{c}{(3)} &
\multicolumn{1}{c}{(4)} & \multicolumn{1}{c}{(5)} & \multicolumn{1}{c}{(6)} &
\multicolumn{1}{c}{(7)} & \multicolumn{1}{c}{(8)} \\
\multicolumn{1}{c}{Number} &  &
\multicolumn{1}{c}{$\alpha$} &
\multicolumn{1}{c}{$\delta$} & \multicolumn{1}{c}{R} &
\multicolumn{1}{c}{S$_{1.4}$} &
\multicolumn{1}{c}{S$_{1.7}$} & \multicolumn{1}{c}{$\alpha$} \\
                              &  & \multicolumn{1}{c}{(J2000)} &
\multicolumn{1}{c}{(J2000)} & \multicolumn{1}{c}{(\arcsec)} &
\multicolumn{1}{c}{(\microJy)} &
\multicolumn{1}{c}{(\microJy)} &  \\
\hline

1  &  & $00:23:01.22\pm$7  & $-72:04:39.4\pm$3	& 298 & 448 & 382 & $-0.8\pm$0.8 \\
2  &  & $00:23:35.60\pm$13 & $-72:01:25.9\pm$6	& 248 & 263 & 165 & $-2.4\pm$2.0 \\
3  &  & $00:23:40.78\pm$4  & $-72:05:11.7\pm$2	& 117 & 744 & 688 & $-0.4\pm$0.5 \\
4  &C & $00:23:50.37\pm$11 & $-72:04:31.3\pm$5	& 74  & 309 & 177 & $-2.9\pm$1.3 \\
5  &  & $00:23:59.41\pm$5  & $-72:03:58.5\pm$2	& 61  & 657 & 416 & $-2.4\pm$0.7 \\
6  &  & $00:24:10.28\pm$9  & $-72:07:41.1\pm$4	& 171 & 367 & 287 & $-1.3\pm$1.0 \\
7  &  & $00:24:13.22\pm$15 & $-72:03:32.4\pm$7	& 86  & 226 & 246 & $+0.4\pm$1.5 \\
8  &D & $00:24:13.94\pm$9  & $-72:04:43.7\pm$4	& 38  & 355 & 269 & $-1.4\pm$1.1 \\
9  &  & $00:24:19.37\pm$17 & $-72:05:17.2\pm$8	& 68  & 197 & $<$ 140 & $<-1.8$  \\
10 &  & $00:24:28.75\pm$14 & $-72:00:28.6\pm$6	& 283 & 241 & 227 & $-0.3\pm$1.5 \\
11 &  & $00:24:30.38\pm$16 & $-72:05:06.6\pm$8	& 114 & 201 & $<$ 140 & $<-1.9$  \\

\end{tabular}
\end{table*}

The uncertainties in source positions derived from the numerical model fitting
do not account for all contributing effects and are typically too small by a
factor of ten. Two alternate methods were used to estimate the uncertainties in
source position. Several sources are detected in several of the five sub-images
which allowed the dispersion of measured positions to be determined directly as
their standard deviation $\Delta\theta_{m}$. This quantity was compared with
$\Delta\theta_{c} \simeq FWHM/snr$ where $\mbox{FWHM} \simeq 5\arcsec$ is the
synthesized beam size and the signal-to-noise ratio $snr = S_{1.4}/\sigma
\simeq S_{1.4}/42$\microJy. For the sources with a number of independent
position measurements, $\Delta\theta_{m}$ and $\Delta\theta_{c}$, differ by
less than 20\% and so $\Delta\theta_{c}$ was used for all sources. The
uncertainties in right ascension and declination are assumed independent and
quoted in Table 2 as $\Delta\alpha = \sqrt{2}\Delta\theta_{c}$ and
$\Delta\delta = \sqrt{2}\Delta\theta_{c}$.

\subsection{Source distribution}
From Fig. 1 it is evident that the radio sources are clustered near the image
centre. Our analysis of source positions is illustrated in Fig. 2 which shows
the cumulative source count as a function of distance from the cluster centre.
To quantify the size of the central group, the radial density variation N(r)
has been modelled by a gaussian function added to a uniformly distributed
background distribution

\[
N(r) = a e^{-(r/b)^2} + c
\]

The fit given by $a = 0.96 \mbox{ arcmin}^{-2}$ , $b = 91\arcsec$, $c = 0.035
\mbox{ arcmin}^{-2}$ is plotted on Fig. 2.  The value of the parameter $c$ (which
corresponds to 2.7 sources in the analysed area) can
be compared with the expected density of background sources.  From an analysis
of the Phoenix Deep Survey \cite{hmcr98}, taking account for extended sources
falling below the brightness detection limit and assuming a Poisson
distribution of source counts in a given area of sky, we find an expected
value for $c$ of $0.10\pm0.04 \mbox{ arcmin}^{-2}$ or $7.8\pm2.8$ sources over
the 10\arcmin\ field.  The number of sources detected in the field is similar to
the expected number of background sources.  Note however that 7 of the
observed sources lie within 2\arcmin\ of the cluster centre, compared with the
$1.3\pm1.1$ sources expected from the Phoenix Deep Survey source counts.

\begin{figure}
\psfig{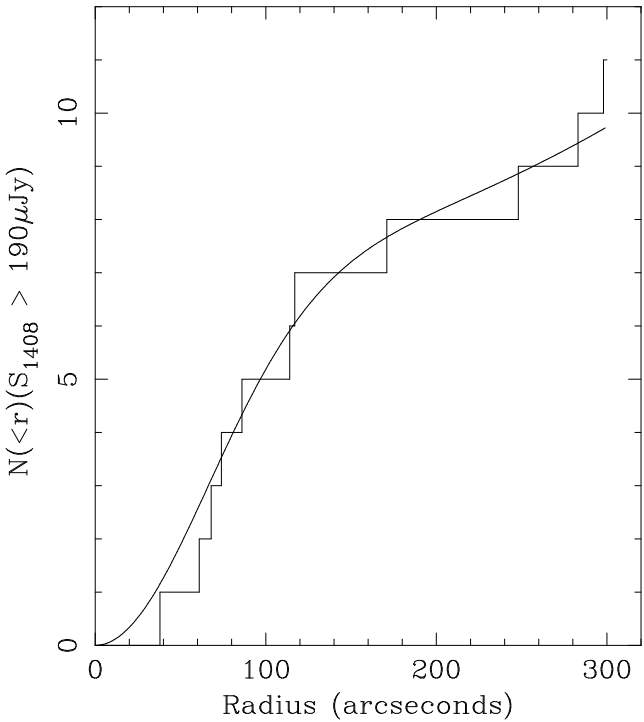}
\caption{Cumulative count of source number inside radius $r$. The curved line
indicates the cumulative source count from model density distribution $N(r) =
ae^{-(r/b)^{2}} + c$ with $a = 0.96\mbox{ arcmin}^{-2}$, $b = 91\arcsec$, $c =
0.035$ arcmin$^{-2}$.}
\end{figure}

\subsection{Source variability}
Observations of the pulsed emission from the pulsars in 47 Tuc \cite{rlm+95} show
their flux density to vary strongly, presumably due to interstellar
scintillation. However, most extragalactic radio sources have a larger angular
extent (while still appearing point-like with the ATCA) and so do not
scintillate, or do so only weakly. The observations described here allow us to
make several flux measurements of some of the 47 Tuc sources. Each
sub-image was examined at every source location listed in Table 2. A source was
deemed visible in a sub-image if its flux exceeded 3 times the image rms from
Table 1. Seven of the 11 sources had measureable flux by this criterion in three or more of the sub-images.
Their fluxes are plotted in Fig. 3.

\begin{figure}
\psfig{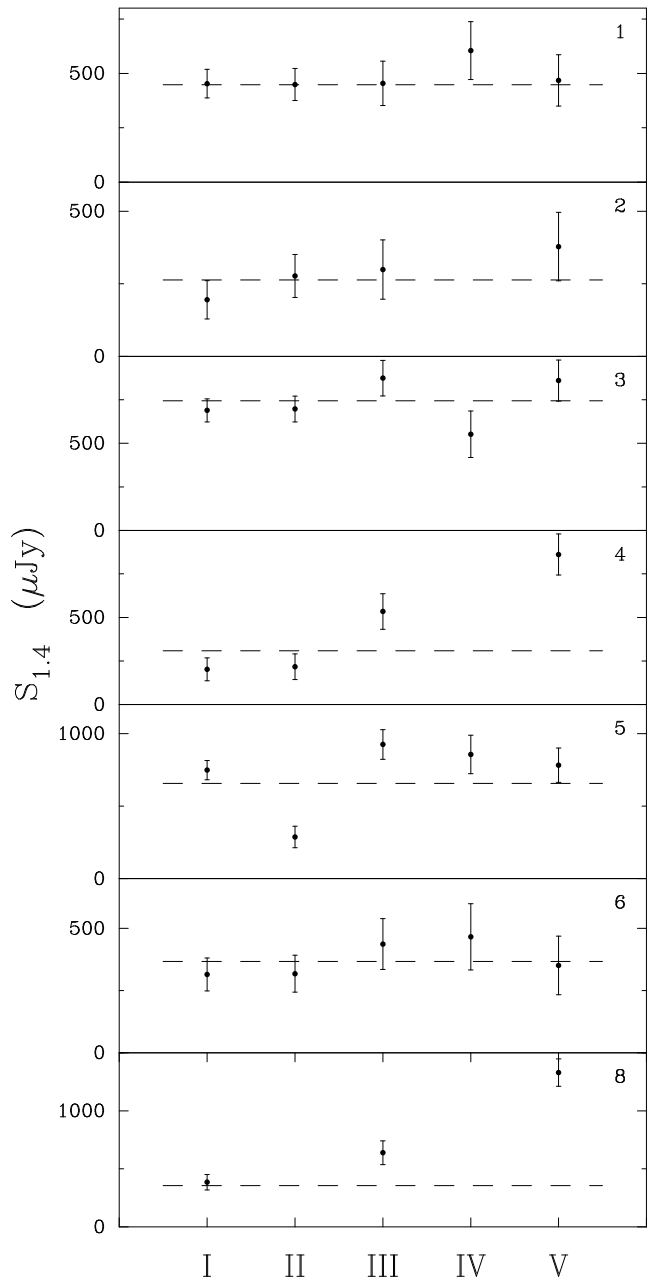}
\caption{1.4 GHz fluxes of a selection of sources (identified with entries in
Table 2 by the number in the upper right) in the field. Up to five flux values
are plotted for each source, corresponding to the five independent observations
listed in Table 1. The horizontal broken line indicates the source flux
measured in the final mosaiced image. The error bars on each point indicate the
rms of pixel brightness in each case.}
\end{figure}

\section{Discussion}

We have presented in Table 2 a list of 11 radio sources within 5\arcmin\ of
the centre of 47 Tuc. The positions of two of these, 4 and 8, can be
identified with pulsars 47 Tuc C and D respectively \cite{rlm+95}.
These three sources have steep spectra, $\alpha_{4} = -2.9$ and $\alpha_{8} =
-1.4$ and varying flux denstities as indicated in Fig. 3. Robinson et. al.
report the spectral indices of 47 Tuc C (0.0) and D (-1.4) measured between
436~MHz and 640~MHz, suggesting that 47 Tuc D has a straight spectrum with
slope -1.4 from 436~MHz to 1.7~GHz but that the spectrum of 47 Tuc C steepens
above 640~MHz.

Source numbers 1 and 10 in Table 2 are likely to be background sources ---
both are distant from the cluster centre and have spectral indices typical of
extragalactic objects.  Source 1 shows little flux variation between the five
observations.  Source 3 also has a flat spectrum and is
probably extragalactic, although some flux variability is suggested.  Sources
2 and 5 both exhibit flux variations, 5 strongly so, and have very steep
spectra and so it seems likely that both are pulsars.  The great distance of
source 2 from the cluster centre (248\arcsec) makes it an interesting pulsar
candidate.  Source 5 has a steep spectrum and variability characteristic of a
pulsar and is stronger than sources 4 and 8 (pulsars C and D), but the pulse
timing analyses of Robinson et al. (1995) did not yield pulsar parameters for its
location.  Perhaps source 5 is a pulsar with properties that make the timing analysis
difficult or that cause its pulses to be undetectable, such as a pulse period outside
the range of the original observations, an extraordinary binary motion or a dense
local environment strongly scattering the pulsed radiation.  The case for the
apparently constant source 6 being a pulsar is
not strong, although it has a spectrum a little steeper than expected for
background objects.  Sources 7, 9 and 11 are faint, making variability and
spectral shape difficult to measure.  We note a similarity in the pattern of
intensity variations between sources 4 and 8 (Fig. 3).  We can find no instrumental
effect which could lead to such apparent variations that would not similarly effect
the other sources in the field.  On the other hand, variability induced by
scintillation in sources as separated as these is difficult to understand.
 
Verbunt and Hasinger (1998) report the positions of a number of X-ray sources
in 47 Tuc, nine of which lie close to the cluster core. Fig. 4 shows their
locations relative to the radio source positions and the locations of pulsars
C and D. It is evident that the X-ray sources belong to two populations, those
within the cluster core, and those represented by X4, X6, X11 and X13 which
have a radial distribution similar to that of the radio sources. Verbunt and
Hasinger are unable to provide a positive identification for any of them, but
point out that their luminosities are consistent with them being soft X-ray
transients, cataclysmic variables or recycled pulsars.   The first epoch of the
X-ray detections reported by Verbunt and Hasinger is close to the date of our
most sensitive observations (1992 April).  We note that none of the
X-ray sources coincide with any of the radio sources.

\begin{figure}
\psfig{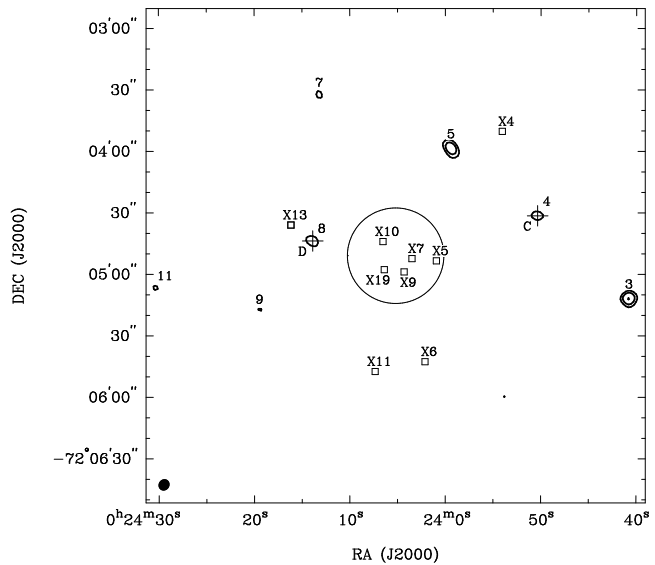}
\caption{The inner part of the imaged field. The radio emission is shown as the
1.4~GHz brightness contours at 180, 350 and 700 \microJy/beam, and the sources
identified by their sequence number from Table 2. The size of the synthesized
beam at 1.4~GHz is indicated at lower left. The cluster core is marked as the
open circle in the centre. Locations of the nine X-ray sources reported by
Verbunt and Hasinger (1998) are shown as open squares and labelled with numbers
from Vebunt and Hasinger's Table 2.}
\end{figure}

We have presented a radio image of the central region of 47 Tuc and shown a
concentration of radio sources close to the cluster core.  Although a direct
correspondence between radio sources and pulsars is possible in only two
cases, it seems likely that several of the other detected sources are also
pulsars. Thus the distribution of sources in the image is likely to be
representative of the spatial distribution of pulsars in the cluster.  We can
now differentiate several classes of object, each with its own characteristic
radial distribution.  There is a class of X-ray object represented by sources
X5, X7, X9, X10 and X19 in the catalogue of Verbunt and Hasinger (1998) which
are confined to the stellar core of the cluster.  There is a second class of
X-ray object (represented by X4, X6, X11, X13, Verbunt and Hasinger) which lie
typically 2 -- 5 core radii from the cluster centre.  Finally there are the
radio sources (possibly pulsars) whose distribution is similar to that of the
outer X-ray class, but which may have some members very far ($\sim$10 core radii)
from the cluster core. 

\section*{Acknowledgements}
The Australia Telescope Compact Array is funded by the Commonwealth of
Australia for operation as a National Facility by CSIRO.  The optical image
in this paper is taken from the Digital Sky Survey.  This image is based on
photographic data obtained using the UK Schmidt Telescope.  The UK Schmidt
Telescope was operated by the Royal Observatory Edinburgh, with funding from
the UK Science and Engineering Research Council, until 1988 June, and
therafter by the Anglo-Australian Observatory.  Original plate material is
copyright of the Royal Observatory Edinburgh and the Anglo-Australian
Observatory.  The plate was processed into the present compressed digital
form with their permission.  The Digital Sky Survey was produced at the
Space Telescope Science Institute under US Government grant NAG W-2166.  We are
grateful to the referee, Professor J. Bell Burnell, for her helpful comments
on the manuscript.

\bibliographystyle{mnras}

\label{lastpage}
\end{document}